\documentclass[twocolumn,showpacs,preprintnumbers,amsmath,amssymb]
{revtex4-1}

\usepackage{color}
\usepackage{graphicx}
\usepackage{dcolumn}
\usepackage{bm}

\begin{document}

\preprint{}

\title{$R$-parity violating supersymmetric Barr-Zee type contributions to the fermion electric dipole moment with weak gauge boson exchange
}

\author{Nodoka Yamanaka}
\affiliation{%
Research Center for Nuclear Physics, Osaka University, Ibaraki, Osaka 567-0047, Japan}%

\author{Toru Sato and Takahiro Kubota}
\affiliation{
Department of Physics, Osaka University, Toyonaka, Osaka 560-0043, Japan}%

\date{\today}

\begin{abstract}
The contribution of the $R$-parity violating trilinear couplings in the supersymmetric model to the fermion electric dipole moment is analyzed at the two-loop level.
We show that in general, the Barr-Zee type contribution to the fermion electric dipole moment with the exchange of $W$ and $Z$ bosons is not small compared to the currently known photon exchange one with $R$-parity violating interactions. 
We will then give new upper bounds on the imaginary parts of $R$-parity violating couplings from the experimental data of the electric dipole moments of the electron and of the neutron. 
The effect due to bilinear $R$-parity violating couplings, which needs to be investigated separately, is not included in our analyses.
\end{abstract}

\maketitle

\section{\label{sec:intro}Introduction}
The standard model (SM) of particle physics, although being very successful in interpreting many experimental data, has difficulty in explaining some phenomena such as the matter abundance of our Universe. 
The SM has therefore to be extended.

There are currently many approaches to search for new physics (NP) beyond the SM.
Among many others, the measurement of the electric dipole moment (EDM) is of particular interest.
The reasons are the following. 
The EDM is an observable sensitive to the violation of the parity and the time-reversal symmetry (or equivalently the $CP$). 
The contribution from the SM is in general very small \cite{smedm}, and this fact makes the EDM a sensitive observable to NP with large $CP$ violation. 
The experimental data available are very accurate for a variety of systems such as the neutron ($d_n < 2.9 \times 10^{-26}e\, {\rm cm}$) \cite{baker}, the $^{205}$Tl ($d_{\rm Tl} < 9 \times 10^{-25}e\, {\rm cm}$) \cite{regan} and the $^{199}$Hg atoms ($d_{\rm Hg} < 3.1 \times 10^{-29}e\, {\rm cm}$) \cite{griffith}, the Ytterbium fluoride (YbF) molecule (which gives bound on the electron EDM: $d_e < 1.05 \times 10^{-27}e\, {\rm cm}$) \cite{hudson}, the muon \cite{muong2}, etc.
The EDM is therefore a very good probe of NP.
A new generation of experiment using storage rings is also in preparation, aiming at the measurements of the EDMs of the muon, proton and deuteron \cite{storage}.

On the theoretical side, the minimal supersymmetric standard model is known to be the leading candidate of NP \cite{mssm}.
A general supersymmetric extension of the SM allows baryon number or lepton  number violating interactions, so we must impose the conservation of {\it R-parity} [$R=(-1)^{3B-L +2s}$] to forbid them. 
This assumption is, however, completely {\it ad hoc}, so the $R$-parity violating (RPV) interactions have to be investigated phenomenologically. 
Until now, many RPV interactions were constrained by high energy experiments, low energy precision tests, and cosmological observations \cite{rpvphenomenology}.
Thanks to many efforts in EDM experiments, many phenomenological analyses of models of multi-Higgs \cite{weinberg,barr-zee,chargedhiggsbarr-zee,leigh}, supersymmetry with \cite{susyedm1-loop,susyedm2-loop,susyedm2-looppilaftsis,susyedmgeneral,pospelovreview,susyedmflavorchange} and 
without \cite{barbieri,godbole,chang,herczeg,cch,choi,SVP,faessler,rpvbarr-zee,rpvsfermionbarr-zee,yamanaka2} 
$R$-parity invariance were done, and many $CP$ phases have been constrained so far.


It has been pointed out in 
\cite{cch, choi} that,  in the most general RPV interactions, the leading contribution 
to the neutron EDM is generated at the one-loop level and that it contains both 
bilinear and trilinear RPV couplings. In the course of showing this,  the authors 
of \cite{choi} made a suitable use of the flavor basis in which only one of the 
four $Y=-1/2$ doublet fields bears vacuum expectation value (VEV)
 \cite{SVP}. In other words 
the direction of the  VEV is singled out as the down type Higgs field $H_{d}$ and  sneutrinos 
are not given VEV. It has also been made clear in \cite{choi} that in the absence of bilinear 
RPV terms the leading contributions to the fermion EDM come from two-loop diagrams. 
This is   in agreement with  the observation that  the Barr-Zee type two-loop level diagrams give the leading 
 effect in the absence of the bilinear terms \cite{godbole,chang, rpvbarr-zee}.
In the present paper we choose the same flavor basis as in \cite{choi} and examine the Barr-Zee type two-loop diagrams arising from the trilinear RPV superpotential.

It was argued in \cite{godbole} that, without the bilinear couplings, the fermion EDM receives the largest part from the photon exchange Barr-Zee type diagram, and that the other Barr-Zee type diagrams with $W$ and $Z$ bosons are subleading, on the basis of the analogy between the $R$-parity violation and models with charged Higgs boson \cite{chargedhiggsbarr-zee,chang}.
We must be careful however to the fact that these two processes cannot be described similarly, since the $SU(2)_L$ gauge structure (chirality structure) of the Yukawa interactions with charged Higgs boson differs from that of the $R$-parity violation.
As the chirality structure inside the fermion loop of the Barr-Zee type diagram is different, the na\"{i}ve analogy no longer works.
Moreover, the RPV Barr-Zee type diagram with $W$ boson exchange changes the flavors of the fields in the loop, and consequently many additional combinations of RPV couplings with new flavor structure may be constrained from the EDM experimental data.

The main purpose of this paper is then to evaluate the two-loop level Barr-Zee type diagram with $W$ and $Z$ boson exchange within $R$-parity violation, and to update the constraints on the RPV interactions provided by the current experimental data of the EDMs of the YbF molecule, the neutron and the $^{199}$Hg atom.

This paper is organized as follows.
We first present the RPV  interaction in the next section.
We then present and formulate in Section \ref{sec:zboson} the two-loop level Barr-Zee type diagram with $Z$ boson exchange.
In Section \ref{sec:wboson}, we derive the formula for the RPV Barr-Zee type contribution with $W$ boson exchange.
This will be done in two steps, first by constructing the inner fermion loop.
We then attach this loop to the external fermion line.
In Section \ref{sec:analysis}, we analyze the limits provided by the current experimental data of the YbF molecule, the neutron and the $^{199}$Hg atom.
We finally summarize our work in the last section.

\section{\label{sec:RPV}RPV contribution}

The superpotential of the RPV interactions relevant in this discussion can be written as follows:
\begin{eqnarray}
W_{{\rm R}\hspace{-.5em}/} &=& \frac{1}{2} \lambda_{ijk} \epsilon_{ab}
 L_i^a L_j^b (E^c)_k
 +\lambda'_{ijk} \epsilon_{ab} L_i^a Q_j^b ( D^c)_k \nonumber\\
&& + \frac{1}{2} \lambda''_{ijk} (U^c)_i (D^c)_j (D^c)_k \ ,
\label{eq:superpotential}
\end{eqnarray}
with $i,j,k=1,2,3$ indicating the generation, $a,b=1,2$ the $SU(2)_L$
indices. The $SU(3)_c$ indices have been omitted. $L$ and $E^c$ denote the lepton doublet and singlet left-chiral superfields. $Q$, $U^c$ and $D^c$ denote respectively the quark doublet, up quark singlet and down quark singlet left-chiral superfields. 
As mentioned in Section \ref{sec:intro} the bilinear terms have been omitted legitimately in our discussion. We also neglected the soft breaking terms in the RPV sector.
Also baryon number violating RPV interactions ($\lambda''_{ijk}$) will be omitted from now, to avoid rapid proton decay.
This RPV superpotential gives the following lepton number violating Yukawa interactions:
\begin{eqnarray}
{\cal L }_{ R\hspace{-.5em}/\,} &=&
- \frac{1}{2} \lambda_{ijk} \left[
\tilde \nu_i \bar e_k P_L e_j +\tilde e_{Lj} \bar e_k P_L \nu_i + \tilde e_{Rk}^\dagger \bar \nu_i^c P_L e_j \right.\nonumber\\
&&\hspace{4.5em} -(i \leftrightarrow j ) \Bigr] \nonumber\\
&&-\lambda'_{ijk} \left[
\tilde \nu_i \bar d_k P_L d_j + \tilde d_{Lj} \bar d_k P_L \nu_i +\tilde d_{Rk}^\dagger \bar \nu_i^c P_L d_j \right. \nonumber\\
&&\hspace{3.5em} \left. -\tilde e_{Li} \bar d_k P_L u_j - \tilde u_{Lj} \bar d_k P_L e_i - \tilde d_{Rk}^\dagger \bar e_i^c P_L u_j \right] \nonumber\\
&&
- \frac{1}{2} \lambda_{ijk} \left[ 
( m_{e_j} \tilde \nu_i \, \tilde e_{Rj} - m_{e_i} \tilde \nu_j \, \tilde e_{Ri})\, \tilde e_{Rk}^\dagger
\right.
\nonumber\\
&& \hspace{5em}
\left.
+m_{e_k} (\tilde \nu_i \, \tilde e_{Lj} - \tilde \nu_j \, \tilde e_{Li})\, \tilde e_{Lk}^\dagger
\right] \nonumber\\
&&
- \lambda'_{ijk} \left[ 
-m_{u_j} \tilde e_{Li} \tilde d_{Rk}^\dagger \tilde u_{Rj}
-m_{e_i} \tilde u_{Lj} \tilde d_{Rk}^\dagger \tilde e_{Ri}
\right.
\nonumber\\
&& \hspace{3em}
\left.
+m_{d_j} \tilde \nu_i \, \tilde d_{Rk}^\dagger \tilde d_{Rj}
+m_{d_k} (\tilde \nu_i \, \tilde d_{Lj} - \tilde e_{Li}\, \tilde u_{Lj} )\, \tilde d_{Lk}^\dagger
\right] 
\nonumber\\
&&
+ ({\rm H.c.}) 
 \ .
\end{eqnarray}
The projection of the chirality is given by $P_L \equiv \frac{1}{2}(1-\gamma_5)$ [and we also define $P_R \equiv \frac{1}{2}(1+\gamma_5)$ for later use].
The RPV scalar 3-point interactions were obtained by combining the RPV superpotential (\ref{eq:superpotential}) with the usual Higgs-matter superpotential.
They are needed to construct the Barr-Zee type diagram with sfermion inner loop \cite{rpvsfermionbarr-zee}.

\section{\label{sec:zboson}Barr-Zee type diagram with $Z$ boson exchange}

\begin{figure}[htb]
\includegraphics[width=6.4cm]{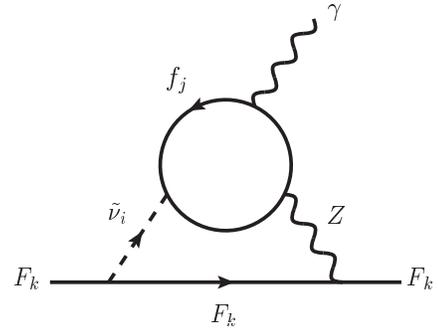}
\caption{\label{fig:Z_barr-zee} 
An example of Barr-Zee type diagram with $Z$ boson exchange generated with RPV interactions.
}
\end{figure}

The RPV interactions contribute to the fermion EDM starting from the two-loop level.
In this section, we will give the formula for the Barr-Zee type diagram with $Z$ boson exchange (see Fig. \ref{fig:Z_barr-zee}).
The computation of this diagram is very similar to that of the Barr-Zee type diagram with photon exchange \cite{rpvbarr-zee,rpvsfermionbarr-zee}.

\begin{figure}[htb]
\includegraphics[width=9cm]{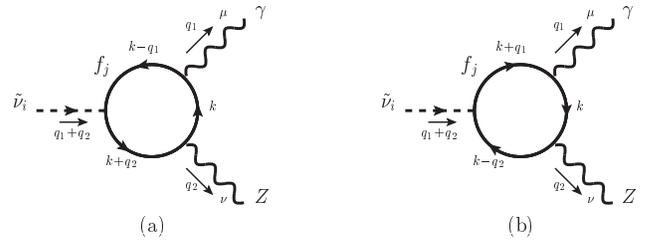}
\caption{\label{fig:nugz}
Fermion loop effective $\tilde \nu \gamma Z$ vertex generated with RPV interactions.
Diagrams (a) and (b) are fermion loop diagrams for the two independent ordering of the photon and the $Z$ boson interactions.
}
\end{figure}

\begin{figure}[htb]
\includegraphics[width=9cm]{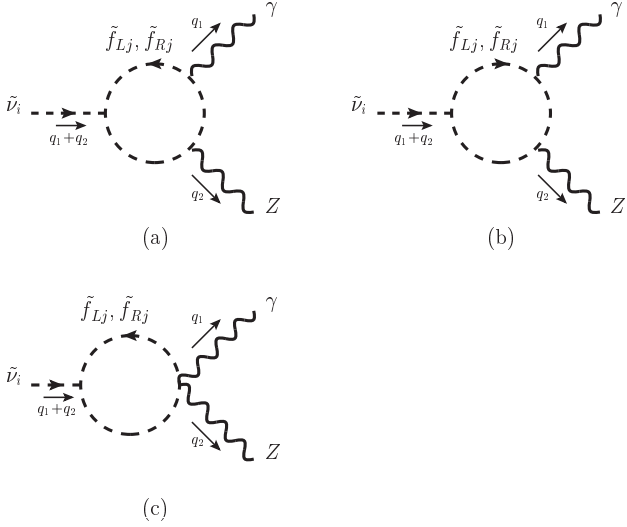}
\caption{\label{fig:nugz_sfermion} 
Sfermion loop effective $\tilde \nu \gamma Z$ vertex generated with RPV interactions.
Diagrams (a) and (b) are the sfermion loop contribution similar to diagrams (a) and (b) of Fig. \ref{fig:nugz} by replacing the fermion by the sfermion.
Diagram (c) newly appears due to the $\gamma-Z$ interaction.
}
\end{figure}

The first step of the evaluation of the Barr-Zee type two-loop diagram is to construct the gauge invariant effective $\tilde \nu \gamma Z$ vertex with the fermion and sfermion loop diagrams, as shown in Figs. \ref{fig:nugz} and \ref{fig:nugz_sfermion}.
This method is based on the analyses of Refs. \cite{leigh,chargedhiggsbarr-zee,chang}.
We will then attach this gauge invariant effective vertex to the external fermion line to obtain the EDM operator.
The amplitude of the inner fermion loop is given as
\begin{widetext}
\begin{eqnarray}
i{\cal M}_{\tilde \nu \gamma Z}&=&
-\hat \lambda_{ijj} n_c Q_f e^2 \alpha_f \epsilon^*_\mu (q_1) \epsilon^*_\nu (q_2) 
\int \frac{d^4 k }{(2\pi)^4} \frac{{\rm Tr} \left[ 
(k\hspace{-.5em} / + q\hspace{-.5em} /_2  +m_{f_j}  ) (1-\gamma_5) (k\hspace{-.5em} / - q\hspace{-.5em} /_1  +m_{f_j}  ) \gamma^\mu (k\hspace{-.5em} /  +m_{f_j}  ) \gamma^\nu \right]}{\left[(k+q_2)^2-m_{f_j}^2 \right] \left[k^2-m_{f_j}^2\right] \left[(k-q_1)^2-m_{f_j}^2 \right]} \nonumber\\
&&
+\sum_{\tilde f_j = \tilde f_{Lj},\tilde f_{Rj}}
2 \hat \lambda_{ijj} n_c m_{f_j} Q_f e^2 \alpha_{\tilde f} \epsilon^*_\mu (q_1) \epsilon^*_\nu (q_2) 
\int \frac{d^4 k }{(2\pi)^4}
\left\{
\frac{(2k^\mu + q_1^\mu ) (2k^\nu - q_2^\nu) }{\left[(k+q_1)^2-m_{\tilde f_j}^2 \right] \left[k^2-m_{\tilde f_j}^2\right] \left[(k-q_2)^2-m_{\tilde f_j}^2 \right]} \right. \nonumber\\
&& \hspace{26em} \left.
-\frac{g^{\mu \nu} }{\left[(k+q_1+q_2)^2-m_{\tilde f_j}^2 \right] \left[k^2-m_{\tilde f_j}^2\right] }
\right\}
\nonumber\\
&=&
\frac{2i}{(4\pi)^2}m_{f_j} \hat \lambda_{ijj} n_c Q_f e^2 \alpha_f \epsilon^*_\mu (q_1) \epsilon^*_\nu (q_2) 
\int^1_0 dx \frac{\left[1-2x(1-x) \right] \left[ q_2^\mu q_1^\nu -(q_1\cdot q_2 ) g^{\mu \nu}  \right]-i\epsilon^{\mu \nu \alpha \beta} (q_1)_\alpha (q_2)_\beta}{m_{f_j}^2-x(1-x)q_2^2 } 
\nonumber\\
&&
+\sum_{\tilde f_j = \tilde f_{Lj},\tilde f_{Rj}}
\frac{i}{(4\pi)^2}\hat \lambda_{ijj} n_c m_{f_j} Q_f e^2 \alpha_{\tilde f} \epsilon^*_\mu (q_1) \epsilon^*_\nu (q_2) 
\int^1_0\ \hspace{-.5em} dx\, \frac{ 2x (1-x) \left[ q_2^\mu q_1^\nu - (q_1 \cdot q_2 )g^{\mu \nu} \right]}{ m_{\tilde f_j}^2 -x(1-x)q_2^2}
+O(q_1^2)
\, ,
\end{eqnarray}
\end{widetext}
where the last equality was given by taking the leading order contribution of the expansion in $q_1$. This approximation is justified since the EDM is the first order coefficient of the multipolar expansion.
Here $i$ and $j$ are the flavor indices of the incident sneutrino and loop fermion $f$ (or loop sfermion $\tilde f$), respectively, and $\hat \lambda$ is the RPV coupling, where $\hat \lambda = \lambda$ for inner loop charged leptons, and $\hat \lambda = \lambda'$ in the case of quarks.
For quark loops, the number of colors is $n_c =3$ (for the lepton loop, $n_c =1$).
The weak coupling of the fermion is given by $\alpha_f$ ($f=l,d$), where $\alpha_l \equiv \frac{1}{4} (3\tan \theta_W -\cot \theta_W) \approx -0.065$ for the coupling of the $Z$ boson with charged leptons, and $\alpha_d \equiv \frac{1}{12} \tan \theta_W -\frac{1}{4}\cot \theta_W \approx -0.42$ for the coupling with down type quarks.
For the sfermion weak coupling, we have $\alpha_{\tilde f_L}= \alpha_f -\beta_f$ and $\alpha_{\tilde f_R}= \alpha_f +\beta_f$, where $\beta_l = \beta_d = \frac{1}{4}(\tan \theta_W + \cot \theta_W)$.
We must note that $\alpha_f$ is the vector coupling of the $Z$ boson with fermions, and that the axial vector coupling ($\beta_f$) does not contribute in the fermion loop.
Note also that the contribution from $\beta_f$ cancels when $m_{\tilde f_Lj}=m_{\tilde f_Rj}$ in the sfermion loop.

The next step is to put the above effective $\tilde \nu \gamma Z$ vertex into the external fermion line to form the EDM operator.
This manipulation is again very similar to that for the photon or gluon exchange Barr-Zee type diagram \cite{rpvbarr-zee,rpvsfermionbarr-zee}, and is given independently of the gauge as follows:
\begin{eqnarray}
i{\cal M}_{Z{\rm BZ}}
&=&
i\, {\rm Im }(\hat \lambda_{ijj} \tilde \lambda_{ikk}^*) \frac{n_c Q_f \alpha_F e \alpha_{\rm em}}{32\pi^3 } 
\nonumber\\
&&
\times
m_{f_j} \epsilon^*_\mu (q_1) \bar u \sigma^{\mu \nu} (q_1)_\nu \gamma_5 u 
\nonumber\\
&&
\times \int_0^1 \hspace{-.5em} dz \Biggl\{
2\alpha_f  I \Bigl(m_Z^2 \, , \,  m_{\tilde \nu_i}^2 \, , \frac{m_{f_j}^2}{z(1-z)} \Bigr)
\nonumber\\
&&\hspace{4em} 
-\hspace{-1.2em}\sum_{\tilde f_j = \tilde f_{Lj},\tilde f_{Rj}} \hspace{-1em} \alpha_{\tilde f}  I \Bigl(m_Z^2 \, , \,  m_{\tilde \nu_i}^2 \, , \frac{m_{\tilde f_j}^2}{z(1-z)} \Bigr)
\Biggr\}
\nonumber\\
&\approx &
i\, {\rm Im }(\hat \lambda_{ijj} \tilde \lambda_{ikk}^*) \frac{n_c Q_f \alpha_f \alpha_F e \alpha_{\rm em}}{16\pi^3 } 
\nonumber\\
&&
\times
\epsilon^*_\mu (q_1) \bar u \sigma^{\mu \nu} (q_1)_\nu \gamma_5 u
\cdot \frac{ m_{f_j} }{m_{\tilde \nu_i}^2 } \ln \frac{m_{\tilde \nu_i}^2}{m_Z^2}\ , 
\label{eq:mzbz}
\end{eqnarray}
where $I(a,b,c)$ is defined as
\begin{eqnarray}
I(a,b,c) &=&
\int_0^\infty \frac{rdr}{(r+a)(r+b)(r+c)} \nonumber\\
&=& 
\frac{1}{(b-a)(c-b)(a-c)}\nonumber\\
&& \times 
 \left[ \ ab \ln \left| \frac{a}{b} \right| + bc \ln \left| \frac{b}{c} \right| +ca \ln \left| \frac{c}{a} \right| \ \right]
\, .
\label{eq:integformula}
\end{eqnarray}
In the last approximation of Eq. (\ref{eq:mzbz}), we have used $m_{\tilde f_j}^2 , m_{\tilde \nu_i}^2 \gg m_Z^2 \gg m_{f_j}^2$.
We have also neglected the sfermion loop contribution (the second term of the integrand of the first equality) which is less than 10\% compared with the fermion loop diagram for $m_{\tilde f_j}^2 , m_{\tilde \nu_i}^2 = O$(TeV).
It is however important to note that the sfermion loop contribution interferes destructively.
This is due to the minus sign brought by the fermion loop contribution \cite{rpvsfermionbarr-zee}.
Note that the fermions $f_j$ and $F_k$ are either down type quarks or charged leptons.
The RPV coupling is $\tilde \lambda = \lambda$ for external charged leptons, and $\tilde \lambda = \lambda'$ for the case of quarks.

The EDM $d_F$ is defined as follows
\begin{equation}
{\cal L}_{\rm EDM} = - \frac{i}{2} d_F \bar \psi_F \gamma_5 \sigma^{\mu \nu} \psi_F F_{\mu \nu} \, , 
\end{equation}
where $F_{\mu \nu}$ is the electromagnetic field strength.
The fermion EDM with $Z$ boson exchange generated by RPV interactions is then given by
\begin{equation}
d_{F_k}^Z \approx
-{\rm Im }(\hat \lambda_{ijj} \tilde \lambda_{ikk}^*) \frac{n_c Q_f \alpha_f \alpha_F e \alpha_{\rm em}}{16\pi^3 } 
\frac{ m_{f_j} }{m_{\tilde \nu_i}^2 } \ln \frac{m_{\tilde \nu_i}^2}{m_Z^2}\ .
\label{eq:(7)}
\end{equation}
Let us compare the $Z$ boson exchange contribution with the photon exchange process.
The photon exchange contribution is given by \cite{rpvbarr-zee}
\begin{equation}
d_{F_k}^\gamma \approx 
{\rm Im } ( \hat \lambda_{ijj} \tilde \lambda^*_{ikk}) \frac{n_c Q_f^2 Q_F e \alpha_{\rm em}}{16\pi^3} \cdot \frac{m_{f_j} }{m_{\tilde \nu_i }^2 }
\left( 2 + \ln \frac{m_{f_j}^2 }{m_{\tilde \nu_i }^2} \right) \, .
\label{eq:(8)}
\end{equation}
We remark that $d_{F_k}^Z$ and $d_{F_k}^\gamma$ of Eqs. (\ref{eq:(7)}) and (\ref{eq:(8)}) have the same sign.
The loop factor (the logarithmic factor) has the same order of magnitude, with $\ln \frac{m_{\tilde \nu_i}^2}{m_Z^2}$ being around one half of $-\left( 2 + \ln \frac{m_{f_j}^2 }{m_{\tilde \nu_i }^2} \right)$ for sneutrino masses of order TeV.
For the lepton EDM with lepton loop, lepton EDM with quark loop and quark EDM with lepton loop, the $Z$ boson exchange contribution is small compared to the photon exchange one, since the weak charge of the lepton is small ($\alpha_l = -0.065$).
For the quark EDM with quark loop however, both diagrams have the same order of magnitude.
This means that an important enhancement of the RPV contribution will occur.
The detailed analysis will be done in Section \ref{sec:analysis}.

\section{\label{sec:wboson}Barr-Zee type diagram with $W$ boson exchange}

The computation of the Barr-Zee type diagram with $W$ boson exchange is similar to that of the Barr-Zee type diagram with charged Higgs exchange, as was done in Ref. \cite{chargedhiggsbarr-zee}.
Here we should give the detail of the derivation.

\subsection{Inner fermion loop}

\begin{figure}[htb]
\includegraphics[width=9cm]{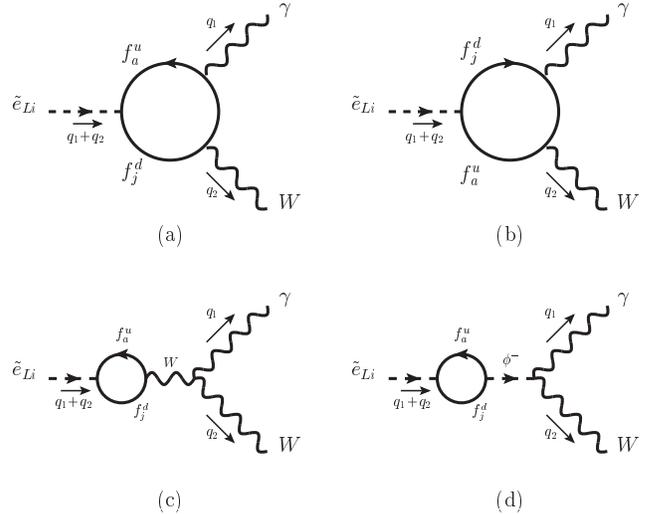}
\caption{\label{fig:egw} 
Fermion loop effective $\tilde e_L \gamma W$ vertex generated with RPV interactions.
Diagrams (a) and (b) are fermion loop diagrams similar to Figs. \ref{fig:nugz} (a) and (b).
Note that for the lepton loop, diagram (a) does not contribute, since the neutrino has no charge.
Diagram (d) is the Nambu-Goldstone boson ($\phi$) exchange contribution, which disappears in the nonlinear $R_\xi$ gauge.
}
\end{figure}

\begin{figure}[htb]
\includegraphics[width=9cm]{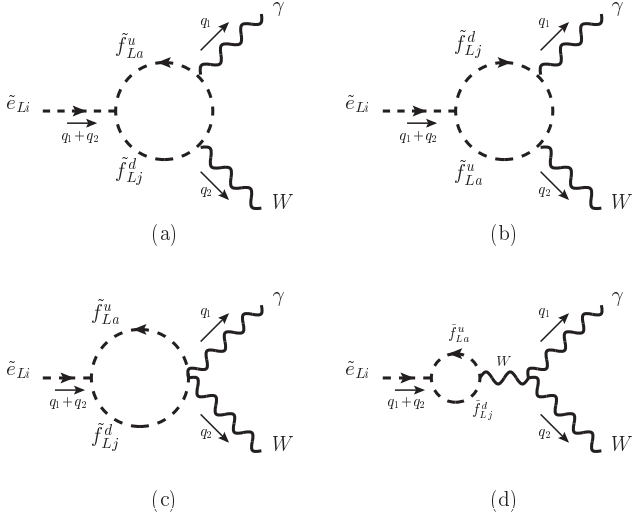}
\caption{\label{fig:egw_sfermion} 
Sfermion loop effective $\tilde e_L \gamma W$ vertex generated with RPV interactions.
Diagrams (a), (b) and (d) are sfermion loop diagrams similar to diagrams (a), (b) and (c) of Fig. \ref{fig:egw}.
Diagram (c) newly appears due to the $\gamma - W$ interaction.
We have omitted the Nambu-Goldstone boson exchange diagram, since it does not contribute in the nonlinear $R_\xi$ gauge.
}
\end{figure}

As for the $Z$ boson exchange contribution, we first derive the gauge invariant expression for the one-loop effective $\tilde e_L \gamma W$ vertex, generated by RPV interactions.
The contributing diagrams are shown in Figs. \ref{fig:egw} and \ref{fig:egw_sfermion}.
In our calculation, we have chosen the nonlinear $R_\xi$ gauge \cite{rxigauge} which is given by the following gauge fixing ($W$ boson)
\begin{equation}
{\cal L}_{GF}^W =
-\frac{1}{\xi } \left| (\partial^\mu -ieA^\mu ) W_\mu^+ -i\xi m_W \phi^+ \right|^2 \ .
\end{equation}
In this gauge, the calculation becomes easy (the interaction between Nambu-Goldstone bosons ($\phi$) and gauge bosons cancels).
The derivation of the one-loop $\tilde e_L \gamma W$ amplitude goes in a manner very similar to that of the decay of the charged Higgs boson into the $W$ boson and photon \cite{hgw,susyedm2-looppilaftsis}.

The amplitude of the one-loop effective $\tilde e_L \gamma W$ vertex with the fermion loop is given by
\begin{equation}
i{\cal M}_{\tilde e \gamma W} = i{\cal M}_{\rm (a)} +i{\cal M}_{\rm (b)} +i{\cal M}_{\rm (c)} 
\ ,
\end{equation}
where 
\begin{widetext}
\begin{eqnarray}
i{\cal M}_{\rm (a)} &=& \hat \lambda_{iaj} \frac{Q_u e^2 V_{aj} }{\sqrt{2} \sin \theta_W} n_c \epsilon^*_\mu (q_1) \epsilon^*_\nu (q_2) \int \frac{d^4 k}{(2\pi )^4} \frac{{\rm Tr}\, \left[ P_L (k \hspace{-.5em}/\, -q\hspace{-.45em}/_1\, + m_{f^u_a}) \gamma^\mu (k\hspace{-.5em}/\, + m_{f^u_a} ) \gamma^\nu P_L (k\hspace{-.5em}/\, + q\hspace{-.5em}/_2\, +m_{f^d_j}) \right]}{\Bigl[(k-q_1 )^2 -m_{f^u_a}^2 \Bigr] \Bigl[k^2 -m_{f^u_a}^2 \Bigr] \Bigl[(k+q_2)^2 -m_{f^d_j}^2 \Bigr]} \ , \label{eq:m(a)}\\
i{\cal M}_{\rm (b)} &=& \hat \lambda_{iaj}  \frac{Q_de^2 V_{aj} }{\sqrt{2} \sin \theta_W} n_c \epsilon^*_\mu (q_1) \epsilon^*_\nu (q_2) \int \frac{d^4 k}{(2\pi )^4} \frac{{\rm Tr}\, \left[ P_L (k \hspace{-.5em}/\, -q\hspace{-.5em}/_2\, + m_{f^u_a}) \gamma^\nu P_L (k\hspace{-.5em}/\, + m_{f^d_j} ) \gamma^\mu (k\hspace{-.5em}/\, + q\hspace{-.5em}/_1\, +m_{f^d_j}) \right]}{\Bigl[(k-q_2 )^2 -m_{f^u_a}^2 \Bigr] \Bigl[k^2 -m_{f^d_j}^2 \Bigr] \Bigl[(k+q_1)^2 -m_{f^d_j}^2 \Bigr]} \ , \label{eq:m(b)} \\
i{\cal M}_{\rm (c)} &=& \hat \lambda_{iaj} \frac{e^2 V_{aj}}{\sqrt{2} \sin \theta_W} n_c \epsilon^*_\mu (q_1) \epsilon^*_\nu (q_2) \int \frac{d^4 k}{(2\pi )^4} 
\frac{{\rm Tr}\, \left[ P_L (k \hspace{-.5em}/\, + m_{f^u_a}) \gamma^{\rho'} P_L (k\hspace{-.5em}/\, +q\hspace{-.5em}/_1 +q\hspace{-.5em}/_2 + m_{f^d_j} )
 \right]}{\Bigl[ k^2 -m_{f^u_a}^2 \Bigr] \Bigl[ (k +q_1 +q_2)^2 -m_{f^d_j}^2 \Bigr] } \nonumber\\
&&\hspace{2em} \times \frac{-1}{(q_1+q_2)^2 -m_W^2} \left[ g^{\rho \rho'} -\frac{(q_1+q_2)^\rho (q_1+q_2)^{\rho'}}{(q_1+q_2)^2 - \xi m_W^2 }(1-\xi) \right]
\nonumber\\
&&\hspace{4em} \times
\left[ \frac{1}{\xi} (q_1^\rho +q_2^\rho )g^{\mu \nu}
+g^{\nu \rho} (q_1^\mu +2q_2^\mu )
+g^{\mu \nu} (q_1^\rho -q_2^\rho )
-g^{\mu \rho} (2q_1^\nu +q_2^\nu )
\right] \ ,
\label{eq:m(c)}
\end{eqnarray}
where $i$, $j$ and $a$ denote respectively the flavor of the incident selectron, down type and up type quarks of the loop.
The convention for $\hat \lambda$ and $\tilde \lambda$ are the same as for the previous case with the Barr-Zee type diagram with $Z$ boson exchange.
The labels $f^u$ and $f^d$ denote respectively the up and down type quarks for quark loop, the neutrino and charged lepton for the lepton loop.
For the quark loop, the number of colors is $n_c = 3$, and $n_c =1$ for the lepton.
Moreover, we have $Q_u = 0$ and $Q_d = -1$ for the lepton loop, and $Q_u = \frac{2}{3}$ and $Q_d = -\frac{1}{3}$ for the quark loop.
Here $V_{aj}$ is the Cabibbo-Kobayashi-Maskawa matrix element for the quark loop contribution.
For the case of lepton loops, $V_{aj}$ is a simple unit matrix, as we do not consider the flavor mixing in the lepton sector. 
(For the realistic case, the neutrinos have small masses $m_\nu$, so the mixing occurs. This effect is however accompanied by a mass insertion, giving a factor of $m_\nu / m_{\rm SUSY}$, so that the mixing contribution can be neglected.)
It should be noted that the diagram (d) in Fig. \ref{fig:egw} vanishes in the nonlinear $R_\xi$ gauge.
The second and third lines of Eq. (\ref{eq:m(c)}) become
\begin{eqnarray}
\frac{-1}{q_{12}^2 -m_W^2} \left[ g^{\rho \rho'} -\frac{q_{12}^\rho q_{12}^{\rho'}}{q_{12}^2 - \xi m_W^2 }(1-\xi) \right]
\Biggl[ 
\frac{1}{\xi} q_{12}^\rho g^{\mu \nu} 
+g^{\nu \rho} (q_1^\mu +2q_2^\mu )
+g^{\mu \nu} (q_1^\rho -q_2^\rho )
-g^{\mu \rho} (2q_1^\nu +q_2^\nu )
\Biggr]
&=&
-\frac{q_{12}^{\rho'}}{q_{12}^2} g^{\mu\nu} 
\ , 
\hspace{2em}
\label{eq:(14)}
\end{eqnarray}
independently of the gauge parameter $\xi$, where $q_{12}=q_1 +q_2$.
The calculation of the one-loop integral can be performed by using the Passarino-Veltman one-loop tensor \cite{passarino-veltman} (see the Appendix for detailed derivation).
By taking the leading gauge invariant contribution first order in $q_1$, we obtain the following amplitude:
\begin{eqnarray}
i{\cal M}_{\tilde e \gamma W} 
&=&
2i \hat \lambda_{iaj} \frac{n_c e^2 V_{aj} }{\sqrt{2} \sin \theta_W} \frac{ m_{f^d_j}}{(4\pi )^2} \epsilon^*_\mu (q_1) \epsilon^*_\nu (q_2) \nonumber\\
&&\hspace{6em} \times \left[ \, Q_u \int_0^1 dz \, \frac{\left[ z(1-z)^2 -z(1-z) \right] \left[ (q_1 \cdot q_2) g^{\mu \nu}-q_2^\mu q_1^\nu \right]
- iz(1-z) \epsilon^{\mu \nu \alpha \beta } (q_1)_\alpha (q_2)_\beta }{ z(1-z)q_2^2  - (1-z)m_{f^u_a}^2 -zm_{f^d_j}^2 } \right. \nonumber\\
&&
\hspace{8em} \left. +Q_d \int_0^1 dz \, \frac{\left[ z^2(1-z) -z^2 \right] \left[ (q_1 \cdot q_2) g^{\mu \nu}-q_2^\mu q_1^\nu \right]
- iz^2 \epsilon^{\mu \nu \alpha \beta } (q_1)_\alpha (q_2)_\beta }{ z(1-z)q_2^2  - (1-z)m_{f^u_a}^2 -zm_{f^d_j}^2 } \, \right] 
+O(q_1^2)
 .\ \ \ 
\label{eq:egwtotfermionloop}
\end{eqnarray}
Diagrams (a), (b) and (c) are each divergent, but the divergence of the total contribution cancels out, as is shown in Eq. (\ref{eq:(A21)}).


The amplitude of the one-loop effective $\tilde e_L \gamma W$ vertex with the sfermion loop is given by
\begin{equation}
i{\cal M}'_{\tilde e \gamma W} = i{\cal M}'_{\rm (a)} +i{\cal M}'_{\rm (b)} +i{\cal M}'_{\rm (c)}+i{\cal M}'_{\rm (d)} 
\ ,
\end{equation}
where 
\begin{eqnarray}
i{\cal M}'_{\rm (a)} &=&
-\hat \lambda_{iaj} V_{aj} n_c m_{f_{j}^d} \frac{Q_u e^2}{\sqrt{2} \sin \theta_W} \epsilon^*_\mu (q_1) \epsilon^*_\nu (q_2) 
\int \frac{d^4 k }{(2\pi)^4}
\frac{(2k^\mu - q_1^\mu ) (2k^\nu + q_2^\nu) }{\left[(k-q_1)^2-m_{\tilde f_{Lj}^u}^2 \right] \left[k^2-m_{\tilde f_{Lj}^u}^2\right] \left[(k+q_2)^2-m_{\tilde f_{Lj}^d}^2 \right]} \, ,
\label{eq:egwsfermion(a)} \\
i{\cal M}'_{\rm (b)} &=&
-\hat \lambda_{iaj} V_{aj} n_c m_{f_{j}^d} \frac{Q_d e^2}{\sqrt{2} \sin \theta_W} \epsilon^*_\mu (q_1) \epsilon^*_\nu (q_2) 
\int \frac{d^4 k }{(2\pi)^4}
\frac{(2k^\mu + q_1^\mu ) (2k^\nu - q_2^\nu) }{\left[(k+q_1)^2-m_{\tilde f_{Lj}^d}^2 \right] \left[k^2-m_{\tilde f_{Lj}^d}^2\right] \left[(k-q_2)^2-m_{\tilde f_{Lj}^u}^2 \right]} \, ,
\\
i{\cal M}'_{\rm (c)} &=&
\hat \lambda_{iaj} V_{aj} n_c m_{f_{j}^d} \frac{(Q_u+Q_d) e^2}{\sqrt{2} \sin \theta_W} \epsilon^*_\mu (q_1) \epsilon^*_\nu (q_2) 
\int \frac{d^4 k }{(2\pi)^4}
\frac{g^{\mu \nu} }{\left[(k+q_1+q_2)^2-m_{\tilde f_{Lj}^d}^2 \right] \left[k^2-m_{\tilde f_{Lj}^u}^2\right] } \, ,
\\
i{\cal M}'_{\rm (d)} &=&
\hat \lambda_{iaj} V_{aj} n_c m_{f_{j}^d} \frac{ e^2}{\sqrt{2} \sin \theta_W} \epsilon^*_\mu (q_1) \epsilon^*_\nu (q_2) 
\int \frac{d^4 k }{(2\pi)^4}
\frac{2k^{\rho'} +q_1^{\rho'} +q_2^{\rho'} }{\left[(k+q_1+q_2)^2-m_{\tilde f_{Lj}^d}^2 \right] \left[k^2-m_{\tilde f_{Lj}^u}^2\right] } \cdot \frac{(q_1+q_2)_{\rho'}}{(q_1+q_2)^2} g^{\mu\nu} 
.
\label{eq:egwsfermion(h)}
\end{eqnarray}
\end{widetext}
Here we have again used Eq. (\ref{eq:(14)}) for the $W$ boson propagator of Eq. (\ref{eq:egwsfermion(h)}).
We obtain then the following amplitude
\begin{eqnarray}
i{\cal M}'_{\tilde e \gamma W}
&=&
-2i \hat \lambda_{iaj} \frac{ n_c e^2 V_{aj}}{\sqrt{2} \sin \theta_W} \frac{ m_{f_{j}^d} }{(4\pi )^2} \epsilon^*_\mu (q_1) \epsilon^*_\nu (q_2) 
\nonumber\\
&&\times
\int_0^1 \hspace{-.5em} dx \,
\frac{Q_u x(1-x)^2 + Q_d x^2 (1-x)}{x(1-x)q_2^2 - (1-x)m_{\tilde f_{La}^u}^2 -xm_{\tilde f_{Lj}^d}^2} 
\nonumber\\
&&\hspace{4em} \times
[ (q_1 \cdot q_2) g^{\mu \nu} -q_2^\mu q_1^\nu ] 
\nonumber\\
&&+O(q_1^2)
\ ,
\label{eq:egwtotsfermionloop}
\end{eqnarray}
where we have written only terms contributing to the EDM.

\subsection{Second loop}

\begin{figure}[htb]
\includegraphics[width=8cm]{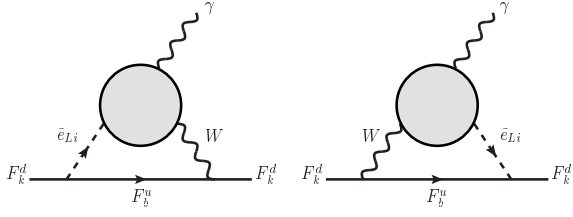}
\caption{\label{fig:W_barr-zee} 
Barr-Zee type contribution to the fermion EDM with $W$ boson exchange generated from RPV interactions.
The diagram on the left side involves the complex conjugated combination of RPV couplings.
$F^u$ and $F^d$ denote respectively the up and down type quarks for quark EDM, the neutrino and charged lepton for lepton EDM.
}
\end{figure}

The Barr-Zee type EDM with $W$ exchange can be constructed by attaching the effective one-loop level $\tilde e_L \gamma W$ vertices generated by the fermion loop $i{\cal M}_{\tilde e \gamma W}$ (\ref{eq:egwtotfermionloop}) and by the sfermion loop $i{\cal M}'_{\tilde e \gamma W}$ (\ref{eq:egwtotsfermionloop}) into the second loop as shown in Fig. \ref{fig:W_barr-zee}.
The RPV Barr-Zee type contribution with $W$ boson exchange and the fermion loop is given by
\begin{widetext}
\begin{eqnarray}
i{\cal M}_{W{\rm BZ}} 
&=&
 \hat \lambda_{iaj} \tilde \lambda_{ibk}^* \frac{n_c e^3 V_{aj} V_{bk} }{ \sin^2 \theta_W} \frac{ m_{f^d_j}}{(4\pi )^2} \epsilon^*_\mu (q_1) \int \frac{d^4 k' }{ (2\pi )^4} \frac{\bar u(p-q_1) \gamma_\nu (p\hspace{-0.45em}/\, -q\hspace{-0.5em}/_1 -k\hspace{-0.5em}/'\, ) P_R u(p)}{\Bigl[ k'^2 - m_W^2 \Bigr] \Bigl[ (k'-p+q_1)^2 - m_{F^u_b}^2  \Bigr] \Bigl[ (k'+q_1)^2 -m_{\tilde e_{Li}}^2 \Bigr]} \nonumber\\
&& \hspace{6em} \times \int_0^1 dz \, \frac{ Q_u (1-z) +Q_d z }{ z(1-z)k'^2  - (1-z)m_{f^u_a}^2 -zm_{f^d_j}^2 } 
\nonumber\\
&& \hspace{10em} \times
\Bigl\{ z(1-z) \left[ (q_1 \cdot k') g^{\mu \nu}-k'^\mu q_1^\nu \right] 
-z \left[ (q_1 \cdot k') g^{\mu \nu}-k'^\mu q_1^\nu
+ i \epsilon^{\mu \nu \alpha \beta } (q_1)_\alpha k'_\beta \right] \Bigr\} 
\nonumber\\
\nonumber\\
&&
- \hat \lambda^*_{iaj} \tilde \lambda_{ibk} \frac{n_c e^3 V_{aj} V_{bk} }{ \sin^2 \theta_W} \frac{ m_{f^d_j}}{(4\pi )^2} \epsilon^*_\mu (q_1) \int \frac{d^4 k' }{ (2\pi )^4} \frac{\bar u(p-q_1) (p\hspace{-0.45em}/\, -q\hspace{-0.5em}/_1 -k\hspace{-0.5em}/'\, ) \gamma_\nu P_L u(p)}{\Bigl[ (k' +q_1)^2 - m_W^2 \Bigr] \Bigl[ (k'-p+q_1)^2 - m_{F^u_b}^2  \Bigr] \Bigl[ k'^2 -m_{\tilde e_{Li}}^2 \Bigr]} \nonumber\\
&& \hspace{6em} \times \int_0^1 dz \, \frac{ Q_u (1-z) +Q_d z }{ z(1-z)k'^2  - (1-z)m_{f^u_a}^2 -zm_{f^d_j}^2 } 
\nonumber\\
&& \hspace{10em} \times
\Bigl\{ z(1-z) \left[ (q_1 \cdot k') g^{\mu \nu}-k'^\mu q_1^\nu \right] 
-z \left[ (q_1 \cdot k') g^{\mu \nu}-k'^\mu q_1^\nu
- i \epsilon^{\mu \nu \alpha \beta } (q_1)_\alpha k'_\beta \right] \Bigr\} 
\nonumber\\
&= &
i\, {\rm Im }(\hat \lambda_{iaj} \tilde \lambda_{ibk}^*) \frac{n_ce \alpha_{\rm em} V_{aj} V_{bk} m_{f^d_j}}{128 \pi^3 \sin^2 \theta_W} \epsilon^*_\mu (q_1) \bar u \sigma^{\mu \nu} (q_1)_\nu \gamma_5 u
\int_0^1 \hspace{-.5em}dz \left[ Q_u (1-z)+Q_d z \right]
I \Bigl(m_W^2 ,  m_{\tilde e_{Li}}^2 , \frac{ m_{f^u_a}^2}{z} +\frac{m_{f^d_j}^2}{1-z} \Bigr)
\nonumber\\
&&+O(q_1^2)
\, ,
\end{eqnarray}
where we have omitted terms proportional to ${\rm Re }(\hat \lambda_{iaj} \tilde \lambda_{ibk}^*)$,
and we have used the formula of Eq. (\ref{eq:integformula}).
We remark that the second term in the curly brackets 
$
-z [ (q_1 \cdot k') g^{\mu \nu}-k'^\mu q_1^\nu
+ i \epsilon^{\mu \nu \alpha \beta } (q_1)_\alpha k'_\beta ]
$
in the $z$ integration vanishes.
This cancellation is reminiscent of the one that we encountered for the RPV Barr-Zee type EDM with the exchange of photons and gluons \cite{rpvbarr-zee}.
Similarly, the sfermion loop contribution is given by
\begin{eqnarray}
i{\cal M}'_{W{\rm BZ}} 
&= &
-i\, {\rm Im }(\hat \lambda_{iaj} \tilde \lambda_{ibk}^*) \frac{n_ce \alpha_{\rm em} V_{aj} V_{bk} m_{f^d_j}}{128\pi^3 \sin^2 \theta_W} \epsilon^*_\mu (q_1) \bar u \sigma^{\mu \nu} (q_1)_\nu \gamma_5 u
\int_0^1 \hspace{-.5em}dz \left[ Q_u (1-z)+Q_d z \right]
I \Bigl(m_W^2 ,  m_{\tilde e_{Li}}^2 , \frac{ m_{\tilde f^u_a}^2}{z} +\frac{m_{\tilde f^d_j}^2}{1-z} \Bigr)
\nonumber\\
&&
+O(q_1^2)
.
\end{eqnarray}
As for the photon \cite{rpvsfermionbarr-zee} and the $Z$ boson exchange contributions, we find that the sfermion loop process interferes destructively with the Barr-Zee type diagram with fermion loop.

We thus obtain the next formula for the EDM with $W$ boson exchange:
\begin{eqnarray}
d_{F^d_k}^W
&= &
-{\rm Im }(\hat \lambda_{iaj} \tilde \lambda_{ibk}^*) \frac{n_ce \alpha_{\rm em} V_{aj} V_{bk} m_{f^d_j}}{128\pi^3 \sin^2 \theta_W} 
\int_0^1 \hspace{-.5em}dz \, (Q_u (1-z)+Q_d z)
\nonumber\\
&&\hspace{16em} \times
\Biggl[
I \Bigl(m_W^2 \, , \,  m_{\tilde e_{Li}}^2 \, , \, \frac{ m_{f^u_a}^2}{z} +\frac{m_{f^d_j}^2}{1-z} \Bigr)  
-I \Bigl(m_W^2 \, , \,  m_{\tilde e_{Li}}^2 \, , \, \frac{ m_{\tilde f^u_a}^2}{z} +\frac{m_{\tilde f^d_j}^2}{1-z} \Bigr)  
\Biggr]
.\ \ \ 
\label{eq:dwinteg}
\end{eqnarray}
\end{widetext}
We treat first the case where there is no top quark in the loop.
In this case, the masses of quarks and leptons can be neglected since $m_{f^u_a}^2 , m_{f^d_j}^2 \ll m_W^2 , m_{\tilde e_{Li}}^2$.
The EDM is then given by
\begin{equation}
d_{F^d_k}^W
\approx
-{\rm Im }(\hat \lambda_{iaj} \tilde \lambda_{ibk}^*) \frac{s_f e \alpha_{\rm em} V_{aj} V_{bk} }{256 \pi^3 \sin^2 \theta_W} 
\frac{m_{f^d_j}}{ m_{\tilde e_{Li}}^2} \ln \frac{m_{\tilde e_{Li}}^2}{m_W^2} \, ,
\label{eq:dwnotop}
\end{equation}
where we have assumed that $m_{\tilde e_{Li}}^2 \gg m_W^2$.
Here $s_f = n_c (Q_u +Q_d)= +1$ for inner quark loop and $s_f = -1$ for lepton loop.
We have verified numerically that the above approximation works well when top quark is absent in the loop.
In this formula, we have also neglected the sfermion loop contribution, since its effect is less than 10\% for sparticle masses of order $O$(1 TeV).

For the case where the top quark is present in the loop, the mass $m_{f^u_a}^2$ cannot be neglected ($ m_{f^d_j}^2 \ll m_W^2 , m_t^2 , m_{\tilde e_{Li}}^2$).
The integral of the fermion loop contribution of Eq. (\ref{eq:dwinteg}) can be performed analytically, and the inner fermion loop contribution to the EDM is given as
\begin{eqnarray}
d_{F^d_k}^W (f)
&\approx &
-{\rm Im }( \lambda'_{iaj} \tilde \lambda_{ibk}^*) \frac{ e \alpha_{\rm em} V_{aj} V_{bk} }{128 \pi^3 \sin^2 \theta_W} \frac{m_{f^d_j}}{m_W^2 - m_{\tilde e_{Li}}^2} \nonumber\\
&&
\times \Biggl[ \left( \frac{3}{z_W^2} - \frac{2}{z_W} \right) \left( {\rm Li}_2 (1-z_W) -\frac{\pi^2}{6} \right)
\nonumber\\
&&\hspace{2em}
+\frac{3}{z_W} \left( 1- \ln (z_W) \right) +\frac{1}{2}\ln (z_W) 
\nonumber\\
&&\hspace{11em}
- (z_W \leftrightarrow z_{\tilde e}) \ \Biggr] ,\ \ \ 
\label{eq:dwtop} 
\end{eqnarray}
where $z_W \equiv m_W^2 / m_t^2$ and $z_{\tilde e} \equiv m_{\tilde e_{Li}}^2 / m_t^2$, and ${\rm Li}_2 (z)$ is the dilogarithm function (for the calculation of the dilogarithm function, we have used the computational code of Ref. \cite{ginsberg}).
The integral with top quark in the loop is of course smaller than the case without it.
For $m_{\tilde e_{Li}} = 1$ TeV, the loop integral with top quarks becomes about 30\% of that without top quarks.
We must note that the sfermion loop contribution is not small in this case.
Numerically, the integral of the sfermion loop is larger than 20\% of the top loop contribution, for $O$(1 TeV) sparticle masses.

\section{\label{sec:analysis}Analysis}

\subsection{Constraints on Im($\hat \lambda_{ijj} \tilde \lambda^*_{ikk}$)}

Let us first discuss the case where no change of generation occurs in the inner fermion/sfermion loop and external fermion line (i.e. $a=j$ and $b=k$).
This contribution interferes with the Barr-Zee type process with photon or gluon exchange, previously investigated in Refs. \cite{rpvphenomenology,godbole,chang,herczeg,faessler,rpvbarr-zee,rpvsfermionbarr-zee}.
This part is thus an extension of these analyses.
In this analysis, we assume that all sparticle masses are equal to 1 TeV \cite{lhc}.
We do not elaborate the details of sparticle mass dependences, although experimental data constraining sparticle masses are accumulating.

The first case to treat is the lepton EDM with inner lepton/slepton loop, where the combinations of RPV couplings ${\rm Im}(\lambda_{311}\lambda^*_{322})$ and ${\rm Im}(\lambda_{211}\lambda^*_{233})$ are involved.
This contribution can be constrained by the current experimental data of the electron EDM given by the YbF molecule experiment \cite{hudson}:
\begin{equation}
|d_e | < 1.05 \times 10^{-27} e \, {\rm cm}\ .
\label{eq:ybf}
\end{equation}
The Barr-Zee type lepton EDM with lepton/slepton loop receives negligible contribution from the $Z$ boson exchange diagram, since the charged leptons have small coupling with the $Z$ boson ($\alpha_e \approx -0.065$).
The $W$ boson exchange effect is also small compared to the photon exchange one by about 1 order of magnitude.
We must also note that the sign of the $W$ boson exchange Barr-Zee type EDM is the same as the photon exchange EDM.
The photon exchange contribution thus dominates, as was claimed in previous works.
The upper limit to the combinations of RPV couplings is shown in Table \ref{table:rpvlimitsfcons}.
To our best knowledge, the experimental data of the electron EDM given by the YbF molecule experiment give the tightest limits.

\begin{table}[htb]
\caption{Upper bounds to the RPV couplings given by the current EDM experimental data of the electron and neutron, with the sparticle mass set to $m_{\rm SUSY}=1$ TeV.
For comparison, we have also written the limits provided only by the photon exchange Barr-Zee type contribution.
Limits given by other experiments \cite{rpvphenomenology,yamanaka2} are also shown
($i=1,2,3$).
}
\begin{ruledtabular}
\begin{tabular}{cccc}
RPV couplings  & This work & $d^\gamma$ only & Other  \\ 
&&&experiments\\
\hline
$|{\rm Im } (\lambda^*_{311} \lambda_{322} )|$ & $2.0 \times 10^{-3}$ & $2.1 \times 10^{-3}$ & 0.15 \\
$|{\rm Im } (\lambda^*_{211} \lambda_{233} )|$ & $1.8 \times 10^{-4}$ & $1.9 \times 10^{-4}$ & 0.25 \\
\hline
$|{\rm Im } (\lambda^*_{211} \lambda'_{211} )|$ & $1.2 \times 10^{-1}$ & $1.0 \times 10^{-1}$ & $7.9 \times 10^{-8}$ \\
$|{\rm Im } (\lambda^*_{311} \lambda'_{311} )|$ & $1.2 \times 10^{-1}$ & $1.0 \times 10^{-1}$ & $7.9 \times 10^{-8}$ \\
$|{\rm Im } (\lambda^*_{211} \lambda'_{222} )|$ & $8.7 \times 10^{-3}$ & $6.7 \times 10^{-3}$ & $3.6 \times 10^{-6}$ \\
$|{\rm Im } (\lambda^*_{311} \lambda'_{322} )|$ & $8.7 \times 10^{-3}$ & $6.7 \times 10^{-3}$ & $3.6 \times 10^{-6}$ \\
$|{\rm Im } (\lambda^*_{211} \lambda'_{233} )|$ & $3.4 \times 10^{-4}$ & $2.9 \times 10^{-4}$ & $1.8 \times 10^{-4}$ \\
$|{\rm Im } (\lambda^*_{311} \lambda'_{333} )|$ & $3.4 \times 10^{-4}$ & $2.9 \times 10^{-4}$ & $1.8 \times 10^{-4}$ \\
\hline
$|{\rm Im } (\lambda^*_{122} \lambda'_{111} )|$ & $3.1 \times 10^{-1}$ & $3.7 \times 10^{-1}$ & $4.4 \times 10^{-2}$ \\
$|{\rm Im } (\lambda^*_{322} \lambda'_{311} )|$ & $3.1 \times 10^{-1}$ & $3.7 \times 10^{-1}$ & $6.0 \times 10^{-3}$ \\
$|{\rm Im } (\lambda^*_{133} \lambda'_{111} )|$ & $2.6 \times 10^{-2}$ & $3.4 \times 10^{-2}$ & $2.6 \times 10^{-2}$ \\
$|{\rm Im } (\lambda^*_{233} \lambda'_{211} )|$ & $2.6 \times 10^{-2}$ & $3.4 \times 10^{-2}$ & $2.9 \times 10^{-2}$ \\
\hline
$|{\rm Im } (\lambda'^*_{i11} \lambda'_{i22} )|$ & 1.4 & 1.2 & $3.1 \times 10^{-4}$ \\
$|{\rm Im } (\lambda'^*_{i11} \lambda'_{i33} )|$ & $3.4 \times 10^{-2}$ & $5.1 \times 10^{-2}$ & $1.1 \times 10^{-5}$ \\
\end{tabular}
\end{ruledtabular}
\label{table:rpvlimitsfcons}
\end{table}

The second case to discuss is the lepton EDM with inner quark/squark loop.
In this case also, the Barr-Zee type diagram with $Z$ boson exchange is small, due to the small weak coupling.
Numerically, it is smaller than 5\%, with the same sign as the photon exchange EDM.
The $W$ boson exchange contribution is however not negligible.
This is because the photon exchange EDM receives a suppression by the fractional charge.
For the strange quark/squark loop contribution, the ratio is $d^W_e / d^\gamma_e \approx 0.22$.
Moreover $d^\gamma_e$ and $d^W_e$ have opposite sign.
This means that the total EDM contribution becomes significantly small.
By using the experimental data of Eq. (\ref{eq:ybf}), it is possible to constrain the combinations of RPV couplings ${\rm Im}(\lambda_{i11}\lambda'^*_{i11})$ $(i=2,3)$, ${\rm Im}(\lambda_{i11}\lambda'^*_{i22})$ $(i=1,3)$ and ${\rm Im}(\lambda_{i11}\lambda'^*_{i33})$ $(i=1,2)$.
For these RPV bilinears, there already exist stronger constraints, given by the experimental data of the $^{199}$Hg atom \cite{griffith} via $P$, $CP$-odd electron-nucleon interaction \cite{herczeg,flambaumdiamagnetic,flambaumformula}.
It is not possible to give new upper limits on the corresponding RPV interactions with Barr-Zee type process.

The third case to consider is the quark EDM with lepton/slepton inner loop.
The $Z$ boson exchange contribution has a small contribution for the same reason as the previous case (numerically, less than 5\%).
The $W$ boson exchange EDM is however sizable, with about 23\% of the photon exchange contribution for the $\mu$ loop contribution, and 35\% for the $\tau$ loop contribution.
The quark EDM with lepton/slepton loop has the same sign as the photon exchange Barr-Zee diagram, so $d^W$ and $d^\gamma$ interfere constructively.
We obtain thus tighter limits for the RPV couplings.
By using the current experimental data of the neutron EDM \cite{baker}
\begin{equation}
|d_n | < 2.9 \times 10^{-26} e \, {\rm cm}\ ,
\label{eq:nedm}
\end{equation}
and the relation between the neutron and quark EDMs calculated with the QCD sum rules \cite{nagata,pospelov}
\begin{equation}
d_n = 0.47 d_d -0.12 d_u +e (0.35 d_d^c +0.17 d_u^c) \, ,
\label{eq:qcdsumrule}
\end{equation}
the combinations of RPV couplings ${\rm Im}(\lambda_{i22}\lambda'^*_{i11})$ $(i=1,3)$ and ${\rm Im}(\lambda_{i33}\lambda'^*_{i11})$ $(i=1,2)$ can be constrained.
The result is given in Table \ref{table:rpvlimitsfcons}.
In this case we could give a new upper limit on ${\rm Im}(\lambda_{i33}\lambda'^*_{i11})$ $(i=1,2)$.

The final case to consider is the quark EDM with inner quark/squark loop.
In this case both $Z$ and $W$ boson exchange contributions are sizable compared with the photon exchange one.
The first reason is that the photon exchange EDM receives a double suppression from the fractional charge of the down type quark.
For the $Z$ boson exchange process, the suppression due to the small weak coupling between $Z$ and charged lepton/slepton is absent so that $d^Z$ and $d^\gamma$ have similar size.
The sign between them is the same and the interference is constructive.
For the $W$ boson, the contribution is around 65\% of the photon exchange EDM for the strange loop, and 33\% for the bottom loop.
We must note that the Barr-Zee type diagram with $W$ boson and inner quark/squark loop acts destructively against the rest.
The RPV interactions contributing to the quark EDM with quark/squark loop can be constrained by the neutron EDM experimental data.
These RPV interactions are, however, already strongly constrained by the experimental data of the $^{199}$Hg atom \cite{griffith} via chromo-EDM and $P$, $CP$-odd 4-quark interactions \cite{pospelov,ban,flambaumdiamagnetic}.
It is not possible to give new upper limits on the corresponding RPV interactions with Barr-Zee type process.

\subsection{Constraints on Im($\hat \lambda_{iaj} \tilde \lambda^*_{ibk}$)}

The Barr-Zee type diagram with $W$ boson exchange provides also possibilities to constrain new combinations of RPV couplings through the generation change of the Cabibbo-Kobayashi-Maskawa matrix.
From the EDM experimental data (\ref{eq:ybf}) and (\ref{eq:nedm}) (with the relation (\ref{eq:qcdsumrule})), we obtain upper limits on the imaginary parts of RPV couplings shown in Table \ref{table:rpvlimitsfch}.

\begin{table*}[htb]
\caption{Upper bounds to the RPV couplings given by the current EDM experimental data with sparticle masses $m_{\rm SUSY}=1$ TeV.
Limits from other experiments \cite{rpvphenomenology,yamanaka2} are also shown.
}
\begin{ruledtabular}
\begin{tabular}{lcc}
RPV couplings  & Limits given by this analysis & Other experiments \\ 
\hline
$|{\rm Im } (\lambda^*_{211} \lambda'_{221} )|$ & 2.6  ($e$ EDM \cite{hudson}) & $2.0 \times 10^{-5}$ \\
$|{\rm Im } (\lambda^*_{311} \lambda'_{321} )|$ & 2.6   ($e$ EDM \cite{hudson}) & $2.0 \times 10^{-5}$ \\
$|{\rm Im } (\lambda^*_{211} \lambda'_{231} )|$ & 260   ($e$ EDM \cite{hudson}) & $8.2 \times 10^{-4}$ \\
$|{\rm Im } (\lambda^*_{311} \lambda'_{331} )|$ & 260   ($e$ EDM \cite{hudson}) & $8.2 \times 10^{-4}$  \\
$|{\rm Im } (\lambda^*_{211} \lambda'_{212} )|$ & 0.13  ($e$ EDM \cite{hudson}) & $3.3 \times 10^{-3}$ \\
$|{\rm Im } (\lambda^*_{311} \lambda'_{312} )|$ & 0.13  ($e$ EDM \cite{hudson}) & $3.3 \times 10^{-3}$ \\
$|{\rm Im } (\lambda^*_{211} \lambda'_{232} )|$ & 2.7  ($e$ EDM \cite{hudson}) & $2.9 \times 10^{-2}$   \\
$|{\rm Im } (\lambda^*_{311} \lambda'_{332} )|$ & 2.7  ($e$ EDM \cite{hudson}) & $2.9 \times 10^{-2}$   \\
$|{\rm Im } (\lambda^*_{211} \lambda'_{213} )|$ & 0.19 ($e$ EDM \cite{hudson}) & $2.9 \times 10^{-2}$ \\
$|{\rm Im } (\lambda^*_{311} \lambda'_{313} )|$ & 0.19 ($e$ EDM \cite{hudson})  & $1.7 \times 10^{-2}$  \\
$|{\rm Im } (\lambda^*_{211} \lambda'_{223} )|$ & $1.6 \times 10^{-2}$ ($e$ EDM \cite{hudson}) & $2.9 \times 10^{-2}$ \\
$|{\rm Im } (\lambda^*_{311} \lambda'_{323} )|$ & $1.6 \times 10^{-2}$ ($e$ EDM \cite{hudson}) & $1.7 \times 10^{-2}$  \\
\hline
$|{\rm Im } (\lambda^*_{122} \lambda'_{121} )|$& 7.0 ($n$ EDM \cite{baker}) & $2.9\times 10^{-2}$ \\
$|{\rm Im } (\lambda^*_{322} \lambda'_{321} )|$& 7.0 ($n$ EDM \cite{baker}) & $2.9\times 10^{-2}$ \\
$|{\rm Im } (\lambda^*_{122} \lambda'_{131} )|$& 180 ($n$ EDM \cite{baker}) & 0.70 \\
$|{\rm Im } (\lambda^*_{322} \lambda'_{331} )|$& 180 ($n$ EDM \cite{baker}) & 0.70 \\
\hline
$|{\rm Im } (\lambda^*_{133} \lambda'_{121} )|$& 0.42 ($n$ EDM \cite{baker}) & $1.7\times 10^{-2}$ \\
$|{\rm Im } (\lambda^*_{233} \lambda'_{221} )|$& 0.42 ($n$ EDM \cite{baker}) & $2.9\times 10^{-2}$ \\
$|{\rm Im } (\lambda^*_{133} \lambda'_{131} )|$& 11 ($n$ EDM \cite{baker}) & 0.42 \\
$|{\rm Im } (\lambda^*_{233} \lambda'_{231} )|$& 11 ($n$ EDM \cite{baker}) & 0.70 \\
\hline
$|{\rm Im}( \lambda'^*_{i 1 1} \lambda'_{i 1 2 })|$ $(i=1,2,3)$ & 7.5 ($n$ EDM \cite{baker}) & $7.3 \times 10^{-4}$ \\
$|{\rm Im}( \lambda'^*_{j 1 1} \lambda'_{j 3 2 })|$ $(j=1,2)$ & 160 ($n$ EDM \cite{baker}) & $8.0 \times 10^{-2}$ \\  
$|{\rm Im}( \lambda'^*_{3 1 1} \lambda'_{3 3 2 })|$ & 160 ($n$ EDM \cite{baker}) &$1.4 \times 10^{-2}$\\ 
$|{\rm Im}( \lambda'^*_{1 2 1} \lambda'_{1 1 2 })|$ & 33 ($n$ EDM \cite{baker}) & $4.0 \times 10^{-4}$ \\
$|{\rm Im}( \lambda'^*_{k 2 1} \lambda'_{k 1 2 })|$ $(k=2,3)$ & 33 ($n$ EDM \cite{baker}) &$3.2 \times 10^{-3}$\\
$|{\rm Im}( \lambda'^*_{i 2 1} \lambda'_{i 2 2 })|$ $(i=1,2,3)$ & 7.6 ($n$ EDM \cite{baker}) & $7.3 \times 10^{-4}$ \\
$|{\rm Im}( \lambda'^*_{i 2 1} \lambda'_{i 3 2 })|$ $(i=1,2,3)$ & 700 ($n$ EDM \cite{baker})& $8.0 \times 10^{-2}$ \\
$|{\rm Im}( \lambda'^*_{i 3 1} \lambda'_{i 1 2 })|$ $(i=1,2,3)$ & 850 ($n$ EDM \cite{baker})&$8.0 \times 10^{-2}$\\
$|{\rm Im}( \lambda'^*_{i 3 1} \lambda'_{i 2 2 })|$ $(i=1,2,3)$ & 200 ($n$ EDM \cite{baker})&$8.0 \times 10^{-2}$\\
$|{\rm Im}( \lambda'^*_{i 3 1} \lambda'_{i 3 2 })|$ $(i=1,2,3)$ & $1.8\times 10^{4}$ ($n$ EDM \cite{baker})& $7.3 \times 10^{-4}$ \\
$|{\rm Im}( \lambda'^*_{i 1 1} \lambda'_{i 1 3 })|$ $(i=1,2,3)$ & 12 ($n$ EDM \cite{baker})&$3.2 \times 10^{-3}$\\
$|{\rm Im}( \lambda'^*_{i 1 1} \lambda'_{i 2 3 })|$ $(i=1,2,3)$ & 0.99 ($n$ EDM \cite{baker})&$3.2 \times 10^{-3}$\\
$|{\rm Im}( \lambda'^*_{i 2 1} \lambda'_{i 1 3 })|$ $(i=1,2,3)$ & 50 ($n$ EDM \cite{baker})&$3.2 \times 10^{-3}$\\
$|{\rm Im}( \lambda'^*_{i 2 1} \lambda'_{i 2 3 })|$ $(i=1,2,3)$ & 4.3 ($n$ EDM \cite{baker})&$3.2 \times 10^{-3}$\\
$|{\rm Im}( \lambda'^*_{1 2 1} \lambda'_{1 3 3 })|$ & 0.67 ($n$ EDM \cite{baker})&$2.0 \times 10^{-4}$\\
$|{\rm Im}( \lambda'^*_{k 2 1} \lambda'_{k 3 3 })|$ $(k=2,3)$ & 0.67 ($n$ EDM \cite{baker})&$8.0 \times 10^{-2}$\\
$|{\rm Im}( \lambda'^*_{1 3 1} \lambda'_{1 1 3 })|$ & $1.3\times 10^{3}$ ($n$ EDM \cite{baker})&$1.7 \times 10^{-5}$\\
$|{\rm Im}( \lambda'^*_{k 3 1} \lambda'_{k 1 3 })|$ $(k=2,3)$ & $1.3\times 10^{3}$ ($n$ EDM \cite{baker})&$8.0 \times 10^{-2}$\\
$|{\rm Im}( \lambda'^*_{i 3 1} \lambda'_{i 2 3 })|$ $(i=1,2,3)$ & 110 ($n$ EDM \cite{baker})&$8.0 \times 10^{-2}$\\
$|{\rm Im}( \lambda'^*_{1 3 1} \lambda'_{1 3 3 })|$ & 18 ($n$ EDM \cite{baker})&$4.9 \times 10^{-3}$\\
$|{\rm Im}( \lambda'^*_{k 3 1} \lambda'_{k 3 3 })|$ $(k=2,3)$ & 18 ($n$ EDM \cite{baker})&2.0\\
\end{tabular}
\end{ruledtabular}
\label{table:rpvlimitsfch}
\end{table*}

We see that the experimental data of the electron EDM give new upper limits on $|{\rm Im } (\lambda^*_{211} \lambda'_{223} )|$ and $|{\rm Im } (\lambda^*_{311} \lambda'_{323} )|$.
These upper bounds could be obtained thanks to the enhancement due to the large bottom quark mass in the inner loop.
Other limits on RPV couplings given by this analysis are weaker than those obtained by other experiments, but we remark that many bounds on RPV couplings are not so different.
The progress on neutron and atomic/molecular EDMs are very promising, and many combinations of RPV couplings are in the reach of the next generation experiments.
The Barr-Zee type EDM with $W$ boson exchange thus provides a very wide accessibility to the $CP$ violation of the RPV sector.

\section{\label{sec:conclusion}Conclusion}

In conclusion we have discussed additional contributions to the fermion EDM in the RPV supersymmetric models.
We have calculated the contribution of the two-loop level Barr-Zee type diagram with $W$ and $Z$ boson exchange.
We have then found that these contributions are not negligible in many situations, contrary to the claim in Ref. \cite{chang}.
In particular, the quark EDM with lepton inner loop was enhanced with the $W$ boson exchange diagram and we have done a significant update of the upper limits to the RPV couplings as $|{\rm Im}( \lambda^*_{j33} \lambda'_{j11})| < 2.6\times 10^{-2}$ ($j=1,2$).
We have also found that many additional RPV contributions generated by the flavor change due to the $W$ boson exchange exist, with many combinations of RPV couplings.
In this case it was also possible to give a new constraint on RPV couplings as $|{\rm Im } (\lambda^*_{j11} \lambda'_{j23} )| < 1.6 \times 10^{-2}$ ($j=2,3$).
Moreover, many RPV couplings can potentially be constrained by near future EDM experiments, since the upper limits given by this analysis and those from other experiments have close values.
The $W$ and $Z$ boson exchange contribution thus provides a very wide possibility to approach the $CP$ violation of the RPV sector.

\appendix

\onecolumngrid

\section{\label{sec:pvint}Passarino-Veltman one-loop integral and the $\tilde e_L \gamma W$ amplitude}

For the calculation of the one-loop $\tilde e_{Li} \gamma W$ process [Eqs. (\ref{eq:m(a)}), (\ref{eq:m(b)}) and (\ref{eq:m(c)})], it is convenient to use the formalism of the Passarino-Veltman one-loop integral \cite{passarino-veltman}.

Let us define the one-loop tensor as
\begin{eqnarray}
B_0 (p_1^2 , m_0^2 , m_1^2 ) &\equiv & \frac{(2\pi \mu)^\epsilon}{i\pi^2} \int d^d k \frac{1}{\bigl[ k^2 - m_0^2 \bigr] \bigl[ (k+p_1)^2 -m_1^2 \bigr]}\, , \\
B^\mu (p_1^2 , m_0^2 , m_1^2 ) &\equiv & \frac{(2\pi \mu)^\epsilon}{i\pi^2} \int d^d k \frac{k^\mu}{\bigl[ k^2 - m_0^2 \bigr] \bigl[ (k+p_1)^2 -m_1^2 \bigr]}\, , \\
C_0 (p_1^2 , (p_1-p_2)^2 , p_2^2 , m_0^2 , m_1^2 , m_2^2 ) &\equiv & \frac{(2\pi \mu)^\epsilon}{i\pi^2} \int d^d k \frac{1}{\bigl[ k^2 - m_0^2 \bigr] \bigl[ (k+p_1)^2 -m_1^2 \bigr] \bigl[ (k+p_2)^2 -m_2^2 \bigr]}\, , \label{eq:c0}\\
C^\mu (p_1^2 , (p_1-p_2)^2 , p_2^2 , m_0^2 , m_1^2 , m_2^2 ) &\equiv & \frac{(2\pi \mu)^\epsilon}{i\pi^2} \int d^d k \frac{k^\mu}{\bigl[ k^2 - m_0^2 \bigr] \bigl[ (k+p_1)^2 -m_1^2 \bigr] \bigl[ (k+p_2)^2 -m_2^2 \bigr]}\, , \label{eq:cmu}\\
C^{\mu \nu} (p_1^2 , (p_1-p_2)^2 , p_2^2 , m_0^2 , m_1^2 , m_2^2 ) &\equiv & \frac{(2\pi \mu)^\epsilon}{i\pi^2} \int d^d k \frac{k^\mu k^\nu}{\bigl[ k^2 - m_0^2 \bigr] \bigl[ (k+p_1)^2 -m_1^2 \bigr] \bigl[ (k+p_2)^2 -m_2^2 \bigr]}\, , \label{eq:cmunu}
\end{eqnarray}
where $d=4-\epsilon$ is the space-time dimension shifted by $\epsilon$.
Because of the Lorentz covariance, $B^\mu$, $C^\mu$ and $C^{\mu \nu}$ can be decomposed as follows
\begin{eqnarray}
B^\mu &=& B_1 p_1^\mu \, ,\\
C^\mu &=& C_1 p_1^\mu +C_2 p_2^\mu\, ,\\
C^{\mu \nu} &=& C_{00}g^{\mu \nu} + C_{11} p_1^\mu p_1^\nu + C_{12} (p_1^\mu p_2^\nu+ p_2^\mu p_1^\nu ) +C_{22}p_2^\mu p_2^\nu \, ,
\end{eqnarray}
where the arguments of the loop functions were omitted.
The loop functions $B_0$, $B_1$, $C_0$, $C_{00}$, $C_{11}$, $C_{12}$ and $C_{22}$ can be explicitly calculated, but they satisfy also many useful relations.
Note also that $B_0$, $B_1$ and $C_{00}$ are divergent.
By contracting $C^\mu$ with external momenta $p_1^\mu$ or $p_2^\mu$, we obtain
\begin{eqnarray}
C_\mu p_1^\mu &=&
\frac{1}{2} B_0 (p_2^2 , m_0^2 , m_2^2)
-\frac{1}{2} B_0 ((p_1-p_2)^2 , m_1^2 , m_2^2) 
-\frac{1}{2} \left[p_1^2 -m_1^2+m_0^2 \right] C_0  \ ,\nonumber\\
C_\mu p_2^\mu 
&=&
\frac{1}{2} B_0 (p_1^2 , m_0^2 , m_1^2)
-\frac{1}{2} B_0 ((p_1-p_2)^2 , m_1^2 , m_2^2)
-\frac{1}{2} \left[p_2^2 -m_2^2+m_0^2 \right] C_0  \ .
\end{eqnarray}
Here we have omitted the arguments for $C_0$ which are implicitly the same as those of Eq. (\ref{eq:c0}).

Similarly, we obtain also additional relations by contracting $C^{\mu \nu}$ with two external momenta:
\begin{eqnarray}
dC_{00} + C_{11} p_1^2 + 2C_{12} (p_1 \cdot p_2) +C_{22} p_2^2 
&=&
B_0 ((p_2 -p_1)^2 , m_1^2 , m_2^2) + m_0^2 C_0 \ , \\
C_{00} +C_{11} p_1^2 + C_{12}(p_1 \cdot p_2)  
&=&
\frac{1}{2} B_1 ((p_2 -p_1)^2 , m_1^2 , m_2^2) 
+\frac{1}{2} B_0 ((p_2 -p_1)^2 , m_1^2 , m_2^2) \nonumber\\
&&-\frac{1}{2} \left[p_1^2 -m_1^2+m_0^2 \right]  C_1 \ ,
\\
C_{12} p_1^2
+ C_{22} (p_1 \cdot p_2) 
&=&
\frac{1}{2} B_1 (p_2^2 , m_0^2 , m_2^2)
-\frac{1}{2} B_1 ((p_2 -p_1)^2 , m_1^2 , m_2^2)\nonumber\\
&&-\frac{1}{2} \left[p_1^2 -m_1^2+m_0^2 \right] C_2 \ ,
\\
C_{11} (p_1\cdot p_2) + C_{12} p_2^2   
&=&
\frac{1}{2} B_1 (p_1^2 , m_0^2 , m_1^2)
+\frac{1}{2} B_1 ((p_2 -p_1)^2 , m_1^2 , m_2^2)\nonumber\\
&&
+\frac{1}{2} B_0 ((p_2 -p_1)^2 , m_1^2 , m_2^2)
-\frac{1}{2} \left[p_2^2 -m_2^2+m_0^2 \right] C_1  
\ , \\
C_{00} +C_{12} (p_1 \cdot p_2) + C_{22} p_2^2 
&=&
-\frac{1}{2} B_1 ((p_2 -p_1)^2 , m_1^2 , m_2^2)
-\frac{1}{2} \left[p_2^2 -m_2^2+m_0^2 \right] C_2 \ ,
\end{eqnarray}
where again arguments for $C$ functions were omitted.
Applying all these relations to Eq. (\ref{eq:m(a)}), we obtain
\begin{eqnarray}
i{\cal M}_{\rm (a)} &=& \frac{2im_{f_j^d}}{(4\pi)^2} \hat \lambda_{iaj} \frac{Q_u e^2 V_{aj} }{\sqrt{2} \sin \theta_W} n_c \epsilon^*_\mu (q_1) \epsilon^*_\nu (q_2) \nonumber\\
&&\times \left\{ \left[((q_1 \cdot q_2) g^{\mu \nu} -q_2^\mu q_1^\nu ) ( 2C_{12} + C_2) +i\epsilon^{\mu \nu \alpha \beta} (q_1)_\alpha (q_2)_\beta C_2 \right] (q_1^2 , (q_1+q_2)^2, q_2^2 , m_{f^u_j}^2 , m_{f^u_j}^2 , m_{f^d_j}^2) \right. \nonumber\\
&&\hspace{2em}\left. +B_1 ((q_1+q_2)^2 , m_{f^u_j}^2 , m_{f^d_j}^2) g^{\mu \nu} \right\}\, .
\label{eq:m(a)appendix}
\end{eqnarray}
Note that the expression in parentheses $(q_1^2 , (q_1+q_2)^2, q_2^2 , m_{f^u_j}^2 , m_{f^u_j}^2 , m_{f^d_j}^2)$ denotes the common arguments for the loop functions $C_{12}$ and $C_2$.
For Eqs. (\ref{eq:m(b)}) and (\ref{eq:m(c)}), we obtain similarly
\begin{eqnarray}
i{\cal M}_{\rm (b)} &=& \frac{2im_{f_j^d}}{(4\pi)^2} \hat \lambda_{iaj}  \frac{Q_de^2 V_{aj} }{\sqrt{2} \sin \theta_W} n_c \epsilon^*_\mu (q_1) \epsilon^*_\nu (q_2)  \nonumber\\
&&\times \left\{ \left[((q_1 \cdot q_2) g^{\mu \nu} -q_2^\mu q_1^\nu ) ( 2C_{12} - C_2- C_0) -i\epsilon^{\mu \nu \alpha \beta} (q_1)_\alpha (q_2)_\beta (C_2 +C_0 ) \right] (q_1^2 , (q_1+q_2)^2, q_2^2 , m_{f^d_j}^2 , m_{f^d_j}^2 , m_{f^u_j}^2) \right.
\label{eq:m(b)appendix}
\nonumber\\
&&\hspace{2em}\left. +\left[ B_1 ((q_1+q_2)^2 , m_{f^d_j}^2 , m_{f^u_j}^2) + B_0 ((q_1+q_2)^2 , m_{f^d_j}^2 , m_{f^u_j}^2) \right] g^{\mu \nu} \right\}, 
\\
i{\cal M}_{\rm (c)} &=& \frac{2im_{f_j^d}}{(4\pi)^2} \hat \lambda_{iaj} \frac{-e^2 V_{aj}}{\sqrt{2} \sin \theta_W} n_c \epsilon^*_\mu (q_1) \epsilon^*_\nu (q_2) g^{\mu \nu} B_1 ((q_1+q_2)^2 , m_{f^u_j}^2 , m_{f^d_j}^2)   \ .
\end{eqnarray}
We should note that the arguments of the $C$ loop functions for $i{\cal M}_{\rm (b)}$ differ from those for $i{\cal M}_{\rm (a)}$.

We should now show that terms with $B$ functions cancel each other for $i{\cal M}_{\rm (a)}+i{\cal M}_{\rm (b)}+i{\cal M}_{\rm (c)}$.
The loop functions $B_0$ and $B_1$ can be directly calculated using Feynman parameters as  
\begin{eqnarray}
B_0 (q^2 , m_0^2 , m_1^2) &=& 
\left[ \frac{2}{\epsilon} -\gamma +\ln (4\pi) \right]
-\int_0^1 dx \, \ln \left( \frac{-x(1-x) q^2 + (1-x)m_0^2 +xm_1^2}{\mu^2} \right) \, , 
\\
B_1 (q^2 , m_0^2 , m_1^2) &=& 
-\frac{1}{2}\left[ \frac{2}{\epsilon} -\gamma +\ln (4\pi) \right]
+\int_0^1 dx \, x\ln \left( \frac{-x(1-x) q^2 + (1-x)m_0^2 +xm_1^2}{\mu^2} \right) \, ,
\end{eqnarray}
where $\gamma$ is the Euler constant and $\mu$ the mass scale peculiar in the dimensional regularization.
By noting that
\begin{eqnarray}
B_0 (q^2 , m_0^2 , m_1^2) +B_1 (q^2 , m_0^2 , m_1^2)
&=& 
\frac{1}{2} \left[ \frac{2}{\epsilon} -\gamma +\ln (4\pi) \right]
-\int_0^1 dx \, (1-x) \ln \left( \frac{-x(1-x) q^2 + (1-x)m_0^2 +xm_1^2}{\mu^2} \right) 
\nonumber\\
&=& 
\frac{1}{2} \left[ \frac{2}{\epsilon} -\gamma +\ln (4\pi) \right]
-\int_0^1 dx \, x \ln \left( \frac{-x(1-x) q^2 + (1-x)m_1^2 +xm_0^2}{\mu^2} \right) 
\nonumber\\
&=&
-B_1 (q^2 , m_1^2 , m_0^2) \ ,
\end{eqnarray}
the sum of the $B$ functions for $i{\cal M}_{\rm (a)}+i{\cal M}_{\rm (b)}+i{\cal M}_{\rm (c)}$ becomes
\begin{eqnarray}
&&Q_u B_1 ((q_1 +q_2)^2 , m_{f^u_j}^2 , m_{f^d_j}^2 ) + Q_d \left[ B_1 ((q_1 +q_2)^2 , m_{f^d_j}^2 , m_{f^u_j}^2 ) + B_0 ((q_1 +q_2)^2 , m_{f^d_j}^2 , m_{f^u_j}^2 )\right] \nonumber\\
&&\hspace{4em}-B_1 ((q_1 +q_2)^2 , m_{f^u_j}^2 , m_{f^d_j}^2 ) \nonumber\\
&=& (Q_u -Q_d -1) B_1 ((q_1 +q_2)^2 , m_{f^u_j}^2 , m_{f^d_j}^2 ) \nonumber\\
&=& 0 \ .
\label{eq:(A21)}
\end{eqnarray}
We see that the divergence cancels.
The evaluation of the remaining $C$ functions can also be done by direct calculation of the integrals.
\begin{eqnarray}
C_{12} (q_1^2 , (q_1+q_2)^2, q_2^2 , m_{f^u_j}^2 , m_{f^u_j}^2 , m_{f^d_j}^2)
&=& 
\int_0^1 \hspace{-.5em}dz \int_0^{1-z} \hspace{-1.2em}dx\,
\frac{xz}{x(1-x)q_1^2 +z(1-z) q_2^2 +2xz (q_1 \cdot q_2) -(1-z) m_{f^u_j}^2 -zm_{f^d_j}^2} \, ,\nonumber\\
\\
C_2 (q_1^2 , (q_1+q_2)^2, q_2^2 , m_{f^u_j}^2 , m_{f^u_j}^2 , m_{f^d_j}^2)
&=& 
\int_0^1 \hspace{-.5em}dz \int_0^{1-z} \hspace{-1.2em}dx\,
\frac{-z}{x(1-x)q_1^2 +z(1-z) q_2^2 +2xz (q_1 \cdot q_2) -(1-z) m_{f^u_j}^2 -zm_{f^d_j}^2} \, ,\nonumber\\
\\
C_{12} (q_1^2 , (q_1+q_2)^2, q_2^2 , m_{f^d_j}^2 , m_{f^d_j}^2 , m_{f^u_j}^2)
&=& 
\int_0^1 \hspace{-.5em}dz \int_0^{1-z} \hspace{-1.2em}dx\,
\frac{xz}{x(1-x)q_1^2 +z(1-z) q_2^2 +2xz (q_1 \cdot q_2) -(1-z) m_{f^d_j}^2 -zm_{f^u_j}^2} \, ,\nonumber\\
\\
C_2 (q_1^2 , (q_1+q_2)^2, q_2^2 , m_{f^d_j}^2 , m_{f^d_j}^2 , m_{f^u_j}^2)
&=& 
\int_0^1 \hspace{-.5em}dz \int_0^{1-z} \hspace{-1.2em}dx\,
\frac{-z}{x(1-x)q_1^2 +z(1-z) q_2^2 +2xz (q_1 \cdot q_2) -(1-z) m_{f^d_j}^2 -zm_{f^u_j}^2} \, ,\nonumber\\
\\
C_0 (q_1^2 , (q_1+q_2)^2, q_2^2 , m_{f^d_j}^2 , m_{f^d_j}^2 , m_{f^u_j}^2)
&=& 
\int_0^1 \hspace{-.5em}dz \int_0^{1-z} \hspace{-1.2em}dx\,
\frac{1}{x(1-x)q_1^2 +z(1-z) q_2^2 +2xz (q_1 \cdot q_2) -(1-z) m_{f^d_j}^2 -zm_{f^u_j}^2} \, .\nonumber\\
\end{eqnarray}
By applying the above formulae to Eqs. (\ref{eq:m(a)appendix}) and (\ref{eq:m(b)appendix}), taking the first order approximation in $q_1$, and integrating by $x$, we obtain Eq. (\ref{eq:egwtotfermionloop}).

\twocolumngrid

\end{document}